\documentclass[aps,prd,twocolumn,nofootinbib,showpacs]{revtex4}

\usepackage{graphicx}

\usepackage[colorlinks=true, linkcolor=blue, citecolor=blue, urlcolor=blue]{hyperref}

\begin{document}

\title{Thrust Distribution in Electron-Positron Annihilation using the Principle of Maximum Conformality}

\author{Sheng-Quan Wang$^{1,2}$}
\email[email:]{sqwang@cqu.edu.cn}

\author{Stanley J. Brodsky$^2$}
\email[email:]{sjbth@slac.stanford.edu}

\author{Xing-Gang Wu$^3$}
\email[email:]{wuxg@cqu.edu.cn}

\author{Leonardo Di Giustino$^{2,4}$}
\email[email:]{leonardo.digiustino@gmail.com}

\address{$^1$Department of Physics, Guizhou Minzu University, Guiyang 550025, P.R. China}
\address{$^2$SLAC National Accelerator Laboratory, Stanford University, Stanford, California 94039, USA}
\address{$^3$Department of Physics, Chongqing University, Chongqing 401331, P.R. China}
\address{$^4$Department of Science and High Technology, University of Insubria, via valleggio 11, I-22100, Como, Italy}

\date{\today}

\begin{abstract}

We present a comprehensive and self-consistent analysis for the thrust distribution by using the Principle of Maximum Conformality (PMC). By absorbing all nonconformal terms into the running coupling using PMC via renormalization group equation, the scale in the running coupling shows the correct physical behavior and the correct number of active flavors is determined. The resulting PMC predictions agree with the precise measurements for both the thrust differential distributions and the thrust mean values. Moreover, we provide a new remarkable way to determine the running of the coupling constant $\alpha_s(Q^2)$ from the measurement of the jet distributions in electron-positron annihilation at a single given value of the center-of-mass energy $\sqrt{s}$.

\end{abstract}

\maketitle

The event shape observables in electron-positron annihilation play a crucial role in understanding Quantum Chromodynamics (QCD). In the last three decades, the event shape observables have been extensively studied experimentally and theoretically. In particular, the event shape observables have been used to precisely determine the coupling constant (see e.g.~\cite{Kluth:2006bw} for a review).

Due to the simple initial leptonic state, the event shapes can be measured with a high precision, especially at $Z^0$ peak~\cite{Heister:2003aj, Abdallah:2003xz, Abbiendi:2004qz, Achard:2004sv, Abe:1994mf}. The precision of experimental measurements calls for an equally precise theoretical prediction for event shapes. The next-to-leading order (NLO) QCD calculations are known since 1980~\cite{Ellis:1980wv, Catani:1996jh}, and the next-to-next-to-leading order (NNLO) calculations have been carried out in Refs.\cite{Gehrmann-DeRidder:2007nzq, Weinzierl:2008iv}. Despite the significant progress made in the last years for both the pQCD calculations~\cite{DelDuca:2016csb} and the resummation of large logarithms (see e.g.~\cite{Abbate:2010xh, Banfi:2014sua}), the main obstruction to achieve an accurate value of $\alpha_s$ is not the lack of precise experimental data but the dominant uncertainties of the theoretical calculations, mainly due to the choice of the renormalization scale $\mu_r$.

It is well known that using the conventional scale setting, the renormalization scale is simply set at the center-of-mass energy $\mu_r=\sqrt{s}$, and the uncertainties are evaluated by varying the scale within an arbitrary range, e.g. $\mu_r\in[\sqrt{s}/2,2\sqrt{s}]$. The event shape distributions using the conventional scale setting do not match the experimental data, and the extracted values of $\alpha_s$ in general deviate from the world average~\cite{Tanabashi:2018oca}.

The conventional procedure of setting the renormalization scale introduces an inherent scheme-and-scale dependence for the pQCD predictions. The scheme dependence of the pQCD violates the fundamental principle of the renormalization group invariance. The conventional procedure gives wrong predictions for the Abelian theory--Quantum Electrodynamics (QED), where the scale of the coupling constant $\alpha$ can be set unambiguously by using the Gell-Mann-Low procedure~\cite{GellMann:1954fq}. The resulting perturbative series is in general factorially divergent at large orders like $n!\beta^n_0\alpha^n_s$--the ``renormalon" problem~\cite{Beneke:1998ui}. It has always been discussed whether the inclusion of higher-order terms would suppress the scale uncertainty; however, by simply varying the scale within a given range of values fixed a priori, the estimation of unknown higher-order terms is unreliable, and one cannot judge whether the poor pQCD convergence is the intrinsic property of pQCD series, or is due to improper choice of scale.

The Principle of Maximum Conformality (PMC)~\cite{Brodsky:2011ta, Brodsky:2012rj, Brodsky:2011ig, Mojaza:2012mf, Brodsky:2013vpa} provides a systematic way to eliminate renormalization scheme-and-scale ambiguities. Since the PMC predictions do not depend on the choice of the renormalization scheme, PMC scale setting satisfies the principles of renormalization group invariance~\cite{Brodsky:2012ms, Wu:2014iba}. The PMC provides the underlying principle for the Brodsky-Lepage-Mackenzie (BLM) method~\cite{Brodsky:1982gc} and reduces in the Abelian limit, $N_C\rightarrow0$~\cite{Brodsky:1997jk}, to the standard Gell-Mann-Low method. The PMC scales are fixed by absorbing the $\beta$ terms that govern the behavior of the running coupling via the renormalization group equation (RGE). The divergent renormalon terms disappear and the convergence of pQCD series can be thus greatly improved.

The thrust ($T$) variable~\cite{Brandt:1964sa, Farhi:1977sg} is one of the most frequently studied event shape observables, which is defined as
\begin{eqnarray}
T=\max\limits_{\vec{n}}\left(\frac{\sum_{i}|\vec{p}_i\cdot\vec{n}|}{\sum_{i}|\vec{p}_i|}\right),
\end{eqnarray}
where the sum runs over all particles in the final state, and the $\vec{p}_i$ denotes the three-momentum of particle $i$. The unit vector $\vec{n}$ is varied to define the thrust direction $\vec{n}_T$ by maximizing the sum on the right-hand side. In general, the range of values is $0\leq (1-T)\leq1/2$, where $(1-T)\rightarrow0$ corresponds to the two back-to-back jets and $(1-T)\rightarrow1/2$ is the spherically symmetric events. For the three-particle events, we have $0\leq(1-T)\leq1/3$~\cite{Dasgupta:2003iq}.

At the center-of-mass energy $\sqrt{s}$, the differential distribution for thrust variable $\tau$ $(\tau=(1-T))$ for renormalization scale $\mu_r=\sqrt{s}\equiv Q$ can be written as
\begin{eqnarray}
\frac{1}{\sigma_0}\frac{d\sigma}{d\tau}&=&A(\tau)\,a_s(Q)+B(\tau)\,a^2_s(Q) + {\cal O}(a^3_s),
\end{eqnarray}
where $a_s(Q)=\alpha_s(Q)/(2\pi)$, $\sigma_0$ is tree-level hadronic cross section. The $A(\tau)$, $B(\tau)$, ... are perturbative coefficients. The experimentally measured thrust distribution is normalized to the total hadronic cross section $\sigma_h$,
\begin{eqnarray}
\frac{1}{\sigma_h}\frac{d\sigma}{d\tau}&=&\bar{A}(\tau)\,a_s(Q)+\bar{B}(\tau)\,a^2_s(Q)+{\cal O}(a^3_s).
\end{eqnarray}
The perturbative coefficients $\bar{A}(\tau)=A(\tau)$, and $\bar{B}(\tau)=B(\tau)-3/2C_FA(\tau)$, ... and their general renormalization scale $\mu_r$ dependence $\bar{A}(\tau,\mu_r)$, $\bar{B}(\tau,\mu_r)$, ... can be restored from the RGE.

The perturbative coefficients can be expressed by the $n_f$-term, e.g., the NLO coefficient $\bar{B}(\tau,\mu_r)=\bar{B}(\tau,\mu_r)_{\rm in}+\bar{B}(\tau,\mu_r)_{n_f}\cdot n_f$. After applying the PMC scale setting, we obtain
\begin{eqnarray}
\frac{1}{\sigma_h}\frac{d\sigma}{d\tau}&=&\bar{A}(\tau)a_s(\mu^{\rm pmc}_r)+\bar{B}(\tau,\mu_r)_{\rm con}a^2_s(\mu^{\rm pmc}_r) + {\cal O}(a^3_s),
\label{eventafterPMC}
\end{eqnarray}
the conformal coefficient can be written as
\begin{eqnarray}
\bar{B}(\tau,\mu_r)_{\rm con}=\frac{11C_A}{4T_R}\bar{B}(\tau,\mu_r)_{n_f}+\bar{B}(\tau,\mu_r)_{\rm in},
\label{evenPMCcon}
\end{eqnarray}
where $C_A=3$, and $T_R$=1/2. The PMC scale is
\begin{eqnarray}
\mu^{\rm pmc}_r=\mu_r\exp\left[\frac{3\bar{B}(\tau,\mu_r)_{n_f}}{4 T_R \bar{A}(\tau)}+{\cal O}(a_s)\right].
\label{evenPMCscale}
\end{eqnarray}

The PMC scale $\mu^{\rm pmc}_r$ is independent of the initial renormalization scale $\mu_r$. Multiplied by the scale-independent conformal coefficient, the resulting PMC prediction eliminates the renormalization scale uncertainty. The PMC scale for the NLO-term is set equal to that of the LO-term in order to preserve the renormalization scheme independence of thrust variable~\cite{Shen:2017pdu}.

We have used the RunDec program~\cite{Chetyrkin:2000yt} to evaluate the $\overline{\rm MS}$ scheme running coupling from $\alpha_s(M_Z)=0.1181$~\cite{Tanabashi:2018oca}. The NLO coefficients are calculated by using the EVENT2~\cite{Catani:1996jh} with a high precision. The NNLO coefficients can be calculated using the EERAD3~\cite{Gehrmann-DeRidder:2007nzq}, and are checked using the results of Ref.\cite{Weinzierl:2008iv}.

We have calculated the thrust differential distributions for the wide range of $\sqrt{s}$, which have been measured at LEP experiments. Another important event shape observable C-parameter has also been calculated. The most precise data are obtained at $\sqrt{s}=M_Z$. In this paper, in order to draw definitive conclusions, we present the thrust differential distributions at $\sqrt{s}=M_Z$.

\begin{figure}[htb]
\centering
\includegraphics[width=0.40\textwidth]{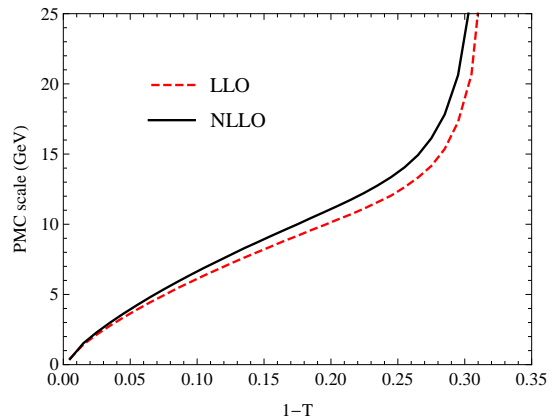}
\caption{The PMC scales up to leading logarithmic order (LLO) and next-leading logarithmic order (NLLO) accuracy for the thrust distribution at $\sqrt{s}=M_Z$. }
\label{figPMCscale}
\end{figure}

The PMC scale is fixed by absorbing the $\beta_i$-terms into the running coupling; it is a perturbative expansion series in $\alpha_s$ and in general shows a fast pQCD convergence. For the present thrust observable, the inclusion of the NNLO correction~\cite{Gehrmann-DeRidder:2007nzq, Weinzierl:2008iv} only slightly changes the PMC scale at NLO level; the PMC scale shows a fast pQCD convergence, as shown explicitly in Fig.(\ref{figPMCscale}). In the following analysis for the thrust distribution and the extraction of $\alpha_s$, we take the PMC scale at NLLO level as determined by using the NNLO correction~\cite{Gehrmann-DeRidder:2007nzq, Weinzierl:2008iv}.

The renormalization scale is simply set at $\mu_r=M_Z$ using the conventional method. The PMC scale is not a single value but it monotonously increases with ($1-T$), reflecting the virtuality of the QCD dynamics. It thus yields the correct physical behavior of the scale and has bound in the two-jet region. Also the number of active flavors $n_f$ changes with $(1-T)$ according to the PMC scale. As the argument of the $\alpha_s$ approaches the two-jet region, the pQCD theory becomes unreliable and the non-perturbative effects must be taken into account. One can adopt the light-front holographic QCD~\cite{Brodsky:2014yha} to evaluate the $\alpha_s$ at the low scale region. In Refs.\cite{Kramer:1990zt, Gehrmann:2014uva}, the correct physical behavior of scale for three-jet process is also obtained. The soft collinear effective theory determines the thrust distribution at different energy scales and also shows that the contribution in two-jet region is affected by the non-perturbative effects~\cite{Abbate:2010xh}.

\begin{figure}[htb]
\centering
\includegraphics[width=0.40\textwidth]{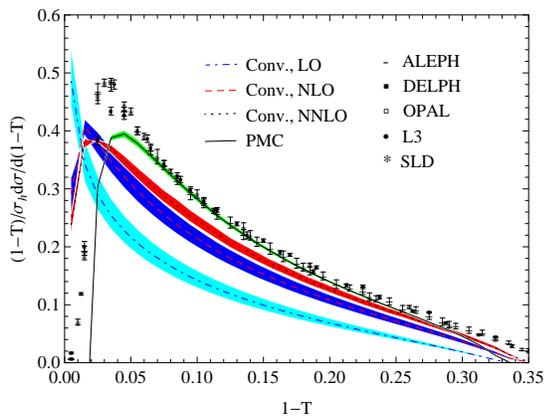}
\caption{The thrust differential distributions using the conventional (Conv.) and PMC scale settings. The dotdashed, dashed and dotted lines are the conventional results at LO, NLO and NNLO~\cite{Gehrmann-DeRidder:2007nzq, Weinzierl:2008iv}, respectively. The solid line is the PMC result. The bands for the theoretical predictions are obtained by varying $\mu_r\in[M_Z/2,2M_Z]$. The PMC prediction eliminates the scale $\mu_r$ uncertainty and its error band is obtained by using $\alpha_s(M_Z)=0.1181\pm0.0011$~\cite{Tanabashi:2018oca}. The experimental data are taken from the ALEPH~\cite{Heister:2003aj}, DELPH~\cite{Abdallah:2003xz}, OPAL~\cite{Abbiendi:2004qz}, L3~\cite{Achard:2004sv} and SLD~\cite{Abe:1994mf} experiments.}
\label{figuseConvT}
\end{figure}

The thrust differential distributions using the conventional and PMC scale settings is shown in Fig.(\ref{figuseConvT}). In the case of the conventional scale setting, we observe that:
\begin{itemize}
\item The NLO and NNLO contributions are always large and positive, except in the two-jet region. The perturbative series for the thrust differential distribution shows a slow convergence.
\item Estimating the magnitude of unknown higher-order QCD corrections by varying the $\mu_r\in[\sqrt{s}/2,2\sqrt{s}]$ is unreliable, i.e., the NLO calculation do not overlap with LO prediction; the NNLO calculation also almost do not overlap with NLO prediction.
\item The conventional predictions are plagued by scale $\mu_r$ uncertainty, and even up to NNLO QCD corrections the conventional predictions do not match the precise experimental data.
\item By fitting the conventional predictions to the experimental data, the extracted coupling constants are deviated from the world average, and are also plagued by significant $\mu_r$ uncertainty~\cite{Dissertori:2007xa}.
\end{itemize}

Due to the kinematical constraints, the domain of the thrust distribution and of the PMC scale at LO is restricted to the range of $0\leq(1-T)\leq1/3$. After applying the PMC, in addition to the small values and the monotonically increasing behavior of the PMC scale, the magnitude of the conformal coefficients are small and its behavior is very different from that of the conventional scale setting. The resulting PMC predictions are in agreement with the experimental data with high precision over the $(1-T)$ region, while they show a slight deviation near the two-jet and multi-jet regions. Based on the conventional scale setting, Ref.\cite{Gehrmann-DeRidder:2007nzq} has also found that due to the presence of large logarithmic contributions and outside of the region of $0.04\leq(1-T)\leq0.33$, the pQCD predictions are unreliable. Thus, in order to improve the predictions near the two-jet and multi-jet regions, the higher pQCD calculations may be needed for the PMC analysis and the resummation of large logarithms should be included. In addition, as we have already mentioned above, the non-perturbative effects should be taken into account in the two-jet region.

In addition to the differential distribution, the mean value of event shapes have also been extensively measured and studied. Since the calculation of the mean value involves an integration over the full phase space, it provides an important platform to complement the differential distribution that afflict the event shapes especially in the two-jet region and to determinate the coupling constant.

The mean value $\langle\tau\rangle$ ($\tau=(1-T)$) of thrust variable is defined by
\begin{eqnarray}
\langle \tau\rangle &=&\int_0^{\tau_0}\frac{\tau}{\sigma_{h}}\frac{d\sigma}{d\tau}d\tau,
\end{eqnarray}
where $\tau_0$ is the kinematical upper limit for the thrust variable.

The electron-positron colliders have collected large numbers of experimental data for the thrust mean value over a wide range of center-of-mass energy (14 GeV $\leq$ $\sqrt{s}$ $\leq$ 206 GeV)~\cite{Heister:2003aj, Abdallah:2003xz, Abbiendi:2004qz, Achard:2004sv, Abe:1994mf, MovillaFernandez:1997fr}. However, the pQCD predictions based on the conventional scale setting substantially deviate from the experimental data. Currently, the most common way is to split the mean value into the perturbative and non-perturbative contributions, which has been studied extensively in the literature. However, some artificial parameters and theoretical models are introduced in order to match the theoretical predictions with the experimental data. It is noted that the analysis of Ref.\cite{Heister:2003aj} obtains a large value of $\alpha_s$ and suggests that a better description for the mean value can be in general obtained by setting the renormalization scale $\mu_r\ll \sqrt{s}$.

The pQCD calculations for the mean value variables have been given in Refs.~\cite{GehrmannDeRidder:2009dp, Weinzierl:2009yz}. After applying the PMC scale setting to the thrust mean value $\langle1-T\rangle$, we obtain the optimal PMC scale,
\begin{eqnarray}
\mu^{\rm pmc}_r|_{\langle1-T\rangle} = 0.0695\sqrt{s},
\end{eqnarray}
which monotonously increases with $\sqrt{s}$, and is 0.0695 times the conventional choice $\mu_r=\sqrt{s}$ and thus $\mu^{\rm pmc}_r|_{\langle1-T\rangle}\ll \sqrt{s}$. We notice that by taking $\sqrt{s}=M_Z=91.1876$ GeV, the PMC scale $\mu^{\rm pmc}_r|_{\langle1-T\rangle}=6.3$ GeV. This is reasonable, since we have shown in Fig.(\ref{figPMCscale}) that the PMC scales of thrust differential distribution are also very small in wide region of ($1-T$). By excluding some results in multi-jet regions, the average of the PMC scale $\langle\mu^{\rm pmc}_r\rangle$ of thrust differential distribution is also close to the $\mu^{\rm pmc}_r|_{\langle1-T\rangle}$. This shows that the PMC scale setting is self-consistent.

\begin{figure}[htb]
\centering
\includegraphics[width=0.40\textwidth]{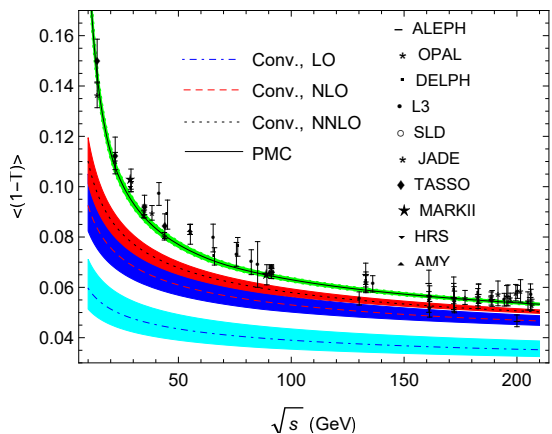}
\caption{Similar to Figure (\ref{figuseConvT}), but for the thrust mean value $\langle 1-T\rangle$ versus the center-of-mass energy $\sqrt{s}$. The conventional result at NNLO is taken from~\cite{GehrmannDeRidder:2009dp, Weinzierl:2009yz}. The measurements are taken from the ALEPH, DELPH, OPAL, L3, SLD, JADE, TASSO, MARKII, HRS and AMY experiments~\cite{Heister:2003aj, Abdallah:2003xz, Abbiendi:2004qz, Achard:2004sv, Abe:1994mf, MovillaFernandez:1997fr}. }
\label{convPMCmoment}
\end{figure}

We present the thrust mean value $\langle 1-T\rangle$ versus the center-of-mass energy $\sqrt{s}$ using the conventional and PMC scale settings in Fig.(\ref{convPMCmoment}). In the case of the conventional scale setting, the perturbative series shows a slow convergence and the estimation of the magnitude of unknown higher-order QCD corrections by varying $\mu_r\in[\sqrt{s}/2,2\sqrt{s}]$ is unreliable. The predictions are plagued by scale $\mu_r$ uncertainty, and substantial deviate from the experimental data even up to NNLO~\cite{GehrmannDeRidder:2009dp}. These cases are similar to those of the thrust differential distributions based on the conventional scale setting.

In contrast, the PMC prediction for the thrust mean value is increased especially in the small center-of-mass energy region. Fig.(\ref{convPMCmoment}) shows that the scale-independent PMC prediction is in excellent agreement with the experimental data in the wide center-of-mass energy range. This suggests that the substantial deviation between the conventional predictions and the experimental data is caused by the improper choice of the renormalization scale. The PMC provides a rigorous explanation for the experimental data without introducing any non-perturbative corrections or artificial parameters.

By taking $\sqrt{s}=M_Z=91.1876$ GeV, in the case of the conventional scale setting, the NLO correction increases the LO prediction by $37\%$ (LO:NLO$\sim0.0395:0.0144$). After applying the PMC, the NLO correction decreases the LO prediction only by $3\%$ (LO:NLO$\sim0.0667:-0.0022$); the PMC prediction at LO is largely increased and the NLO correction is negative and very small. Thus, due to the absorption of the divergent renormalon terms, a strikingly much faster pQCD convergence can be obtained by using the PMC. This leads us to believe that although the higher-order correction is sizeable using conventional scale setting, it will be largely suppressed after using the PMC scale setting.

The thrust distributions have been extensively used to precisely determine the coupling constant. In the case of the conventional scale setting, the extracted $\alpha_s$ indicate a large values compared to the world average. For example, at $\sqrt{s}=M_Z$, a large value $\alpha_s(M_Z)\sim0.1446$~\cite{Dissertori:2007xa} is obtained by fitting the NLO thrust differential distribution with experimental data, which is improved to be $\alpha_s(M_Z)=0.1274\pm0.0047$~\cite{Dissertori:2007xa} (with a perturbative uncertainty of 0.0042) by the inclusion of NNLO correction. The main source of uncertainty for extracted values of $\alpha_s$ is the choice of the renormalization scale. Moreover, the recent determination of $\alpha_s$ by matching the resummation calculations up to N$^3$LL accuracy is $\alpha_s(M_Z)=0.1135\pm0.0011$~\cite{Abbate:2010xh}, which is rather smaller than the PDG world average~\cite{Tanabashi:2018oca}.

We now analyze the extraction of $\alpha_s$ from the thrust differential distribution at $\sqrt{s}=M_Z$ using the PMC. The pQCD calculation corresponds to a parton-level distribution, while the experimental measurements are the hadron-level. Some previous extractions of $\alpha_s$ applied Monte Carlo generators to correct the effects of hadronization. In our present analysis, we adopt the method similar to~\cite{Becher:2008cf} to extract $\alpha_s$.

\begin{figure}[htb]
\centering
\includegraphics[width=0.40\textwidth]{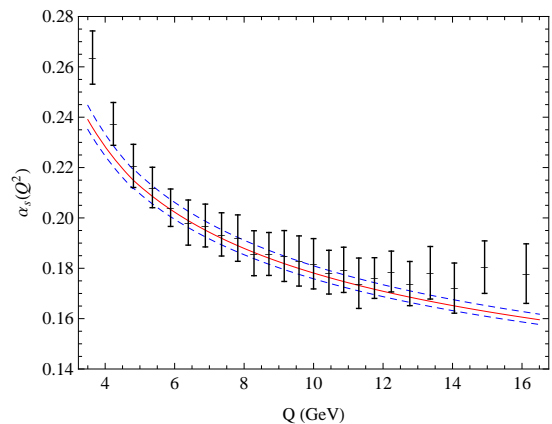}
\caption{The extracted $\alpha_s(Q^2)$ in the $\overline{\rm MS}$ scheme from the comparison of PMC predictions with ALEPH data~\cite{Heister:2003aj} at $\sqrt{s}=M_Z$. The error bars are from the combination of the experimental and theoretical errors. The three lines are the world average evaluated from $\alpha_s(M_Z)=0.1181\pm0.0011$~\cite{Tanabashi:2018oca}. }
\label{figasPMCT}
\end{figure}

A definitive advantage of using the PMC scale setting is that since the PMC scale varies with $(1-T)$, we can extract directly the strong coupling $\alpha_s$ at a wide range of scales using the experimental data at single center-of-mass-energy, $\sqrt{s}=M_Z$. In this case we have used the most precise data from the ALEPH~\cite{Heister:2003aj}. We have calculated the thrust differential distribution at each bin correspondingly to the bins of the experimental data. We can then extract the $\alpha_s$ at different scales bin-by-bin from the comparison of PMC predictions with experimental data. The extracted $\alpha_s$ are explicitly presented in Fig.(\ref{figasPMCT}). It shows that in the scale range of $3.5$ GeV $<Q<16$ GeV (corresponding ($1-T$) range is $0.05<(1-T)<0.29$), the extracted $\alpha_s$ are in excellent agreement with the world average evaluated from $\alpha_s(M_Z)$~\cite{Tanabashi:2018oca}.

In the case of the conventional prescription, the scale is always simply set as $\mu_r=\sqrt{s}=M_Z$, and thus only one value of $\alpha_s$ at scale $M_Z$ can be extracted. In addition, for most of the previous work of extracted $\alpha_s$, the fit range of the thrust ($1-T$) distribution is in general narrow. After using the PMC, we can obtain a self-consistent determination of $\alpha_s$ at different scales over a wide range of the thrust distribution. Moreover, since the PMC predictions eliminate the renormalization scale uncertainty, the extracted $\alpha_s$ are not plagued by any uncertainty in the choice of $\mu_r$. Remarkably, the PMC provides a new way to determine the running of $\alpha_s(Q^2)$ and verify asymptotic freedom from the measurement of the jet distributions in $e^+e^-$ annihilation at a single energy of $\sqrt{s}$.

In conclusion, the thrust variable in $e^+e^-$ annihilation is an ideal platform for testing the QCD. In the case of the conventional scale setting, the predictions are scheme-and-scale dependent and do not match the precise experimental results; the extracted coupling constants in general deviate from the world average. In contrast, after applying PMC scale-setting, we obtain a comprehensive and self-consistent analysis for the thrust variable results including both the differential distributions and the mean values. The PMC scale reflects the virtuality of the QCD dynamics. Moreover, a new remarkable way of extracting $\alpha_s$ at different scales is obtained by comparing the PMC predictions with the experimental data measured at a single center-of-mass-energy $\sqrt{s}$. Our analysis shows the importance of a correct renormalization scale setting, and we expect that the PMC method will be applied to the other event shape variables in electron-electron, electron-proton or proton-proton collisions.

\hspace{1cm}

{\bf Acknowledgements}: We thank Francesco Sannino and Jennifer Rittenhouse West for useful discussions. S.Q.W. thanks the SLAC theory group for kind hospitality and the CSC financial support (No.201708525043). L.D.G. thanks the SLAC theory group for kind hospitality and financial support. This work was supported in part by the Natural Science Foundation of China under Grant No.11625520, No.11705033 and No.11847301; by the Project of Guizhou Provincial Department under Grant No.2016GZ42963, No.KY[2016]028 and No.KY[2017]067; and by the Department of Energy Contract No.DE-AC02-76SF00515. SLAC-PUB-17375.

\end{document}